**Quantum Theory's Reality Problem**

**Adrian Kent**

**Centre for Quantum Information and Foundations, DAMTP, University of Cambridge, U.K.**

**Perimeter Institute for Theoretical of Physics, 31 Caroline Street N, Waterloo, Ontario, Canada**

**Abstract** This review, intended for a popular audience, was originally published in the online magazine Aeon on 28 January 2014. It is reproduced on the arxiv with permission. The online version (without references) can be found at https://aeon.co/essays/what-really-happens-in-schrodinger-s-box.

In 1909, Ernest Rutherford, Hans Geiger and Ernest Marsden took a piece of radium and used it to fire charged particles at a sheet of gold foil. They wanted to test the then-dominant theory that atoms were simply clusters of electrons floating in little seas of positive electrical charge (the so-called 'plum pudding' model). What came next, said Rutherford, was 'the most incredible event that has ever happened to me in my life'.

Despite the airy thinness of the foil, a small fraction of the particles bounced straight back at the source – a result, Rutherford noted, 'as incredible as if you fired a 15-inch shell at a piece of tissue paper and it came back and hit you'. Instead of whooshing straight through the thin soup of electrons that should have been all that hovered in their path, the particles had encountered something solid enough to push back. Something was wrong with matter. Somewhere, reality had departed from the best available model. But where?

The first big insight came from Rutherford himself. He realised that, if the structure of the atom were to permit collisions of the magnitude that his team had observed, its mass must be concentrated in a central nucleus, with electrons whirling around it. Could such a structure be stable? Why didn't the electrons just spiral into the centre, leaking electromagnetic radiation as they fell?

Such concerns prompted the Danish physicist Niels Bohr to formulate a rather oddly rigid model of the atom, using artificial-seeming rules about electron orbits and energy levels to keep everything in order. It was ugly but it seemed to work. Then, in 1924, a French aristocrat and physicist named Louis de Broglie argued [1] that Bohr's model would make more sense if we assumed that the electrons orbiting the atomic nucleus (and indeed everything else that had hitherto been considered a particle) either came with, or in some sense could *behave like*, waves.

If Bohr's atom had seemed a little arbitrary, de Broglie's improved version was almost incomprehensible. Physical theory might have recovered some grip on reality but it seemed to have decisively parted company from common sense. And yet, as Albert Einstein said on reading de Broglie's thesis, here was 'the first feeble ray of light on this worst of our physics enigmas'. By 1926, these disparate intuitions and partial models were already unified into a new mathematical theory called quantum mechanics. Within a few years, the implications for chemistry, spectroscopy and nuclear physics were being confirmed.

It was clear from the start that quantum theory challenged all our previous preconceptions about the nature of matter and how it behaves, and indeed about what science can possibly – even in principle – say about these questions. Over the years, this very slipperiness has made it irresistible to hucksters of various descriptions. I regularly receive ads offering to teach me how to make quantum jumps into alternate universes, tap into my infinite quantum self-energy, and make other exciting-sounding excursions from the plane of reason and meaning. It's worth stressing, then, that the theory itself is both mathematically precise and extremely well confirmed by experiment.

Quantum mechanics has correctly predicted the outcomes of a vast range of investigations, from the scattering of X-rays by crystals to the discovery of the Higgs boson at the Large Hadron Collider. It successfully explains a vast range of natural phenomena, including the structure of atoms and molecules, nuclear fission and fusion, the way light interacts with matter, how stars evolve and shine, and how the elements forming the world around us were originally created.

Yet it puzzled many of its founders, including Einstein and Erwin Schrödinger, and it continues to puzzle physicists today. Einstein in particular never quite accepted it. 'It seems hard to sneak a look at God's cards,' he wrote to a colleague, 'but that he plays dice and uses "telepathic" methods (as the present quantum theory requires of him) is something that I cannot believe for a single moment.' In a 1935 paper [2] co-written with Boris Podolsky and Nathan Rosen, Einstein asked: 'Can [the] Quantum-Mechanical Description of Physical Reality Be Considered Complete?' He concluded that it could not. Given apparently sensible demands on what a description of physical reality must entail, it seemed that *something* must be missing. We needed a deeper theory to understand physical reality fully.

Einstein never found the deeper theory he sought. Indeed, later theoretical work by the Irish physicist John Bell [3] and subsequent experiments suggested that the apparently reasonable demands of that 1935 paper could never be satisfied. Had Einstein lived to see this work, he would surely have agreed that his own search for a deeper theory of reality needed to follow a different path from the one he sketched in 1935.

Even so, I believe that Einstein would have remained convinced that a deeper theory was needed. None of the ways we have so far found of looking at quantum theory are entirely believable. In fact, it's worse than that. To be ruthlessly honest, none of them even quite makes sense. But that might be about to change.

Here's the basic problem. While the mathematics of quantum theory works very well in telling us what to expect at the end of an experiment, it seems peculiarly conceptually confusing when we try to understand what was happening *during* the experiment. To calculate what outcomes we might expect when we fire protons at one another in the Large Hadron Collider, we need to analyse what – at first sight – look like many different stories. The same final set of particles detected after a collision might have been generated by lots of different possible sequences of energy exchanges involving lots of different possible collections of particles. We can't tell which particles were involved from the final set of detected particles.

Now, if the trouble was only that we have a list of possible ways that things could have gone in a given experiment and we can't tell which way they actually went just by looking at the results, that wouldn't be so puzzling. If you find some flowers at your front door and you're not sure which of your friends left them there, you don't start worrying that there are inconsistencies in your understanding of physical reality. You just reason that, of all the people who could have brought them, one of them presumably did. You don't have a logical or conceptual problem, just a patchy record of events.

Quantum theory isn't like this, as far as we presently understand it. We don't get a list of possible explanations for what happened, of which one (although we don't know which) must be the correct one. We get a mathematical recipe that tells us to combine, in an elegant but conceptually mysterious way, numbers attached to each possible explanation. Then we use the result of this calculation to work out the likelihood of any given final result. But here's the twist. Unlike the mathematical theory of probability, this quantum recipe requires us to make different possible stories cancel each other out, or fully or partially reinforce each other. This means that the net chance of an outcome arising from several possible stories can be more or less than the sum of the chances associated with each.

To get a sense of the conceptual mystery we face here, imagine you have three friends, John, Mary and Jo, who absolutely never talk to each other or interact in any other way. If any one of them is in town, there's a one-in-four chance that this person will bring you flowers on any given day. (They're generous and affectionate friends. They're also entirely random and spontaneous – nothing about the particular choice of day affects the chance they might bring you flowers.) But if John and Mary are both in town, you know there's no chance you'll get any flowers that day – even though they never interact, so neither of them should have any idea whether the other one is around. And if Mary and Jo are both in town, you'll certainly get exactly one bunch of flowers – again, even though Mary and Jo never interact either, and you'd have thought that if they're acting independently, your chance of getting any flowers is a bit less than a half, while once in a while you should get two bunches.

If you think this doesn't make any sense, that there has to be something missing from this flower delivery fable, well, that's how many thoughtful physicists feel about quantum theory and our understanding of nature. Pretty precisely analogous things happen in quantum experiments.

One attempt to make sense of this situation – the so-called 'Copenhagen interpretation' of quantum theory, versions of which were advocated by Bohr, Werner Heisenberg and other leading quantum theorists in the first half of the last century – claims that quantum theory is teaching us something profound and final about the limits of what science can tell us. According to this approach, a scientific question makes sense *only if* we have a direct way of verifying the answer. So, asking what we'll see in our particle detectors is a scientific question; asking what happened in the experiment before anything registered in our detectors isn't, because we weren't looking. To be looking, we'd have had to put detectors in the middle of the experiment, and then it would have been a different experiment. In trying to highlight the absurd-seeming consequences of this view, Schrödinger minted what has become its best-known popular icon – an imaginary experiment with a sealed box containing a cat that is simultaneously alive and dead, only resolving into one or other definite state when an experimenter opens the box.

The Copenhagen interpretation was very much in line with the scientific philosophy of logical positivism that caught on at around the same time. In particular, it rests on something like logical positivism's *principle of verification*, according to which a scientific statement is meaningful only if we have some means of verifying its truth. To some of the founders of quantum theory, as well as to later adherents of the Copenhagen interpretation, this came to seem an almost self-evident description of the scientific process. Even after philosophers largely abandoned logical positivism – not least because the principle of verification fails its own test for meaningful statements – many physicists trained in the Copenhagen tradition insisted that their stance was no more than common sense.

However, its consequences are far from commonsensical. If you take this position seriously, then you have to accept that the Higgs boson wasn't actually discovered at the Large Hadron Collider, since no one has ever directly detected a Higgs boson, and we have no direct evidence to support the claim that the Higgs boson is a real particle. Insofar as we learnt anything about nature from the Large Hadron Collider, it was merely what sort of records you get in your detectors when you build something like the Large Hadron Collider. It's hard to imagine the scientists who work on it, or the citizens who funded them, being very enthusiastic about this justification, but on a strict Copenhagen view it's the best we can do.

It gets worse. Quantum theory is supposed to describe the behaviour of elementary particles, atoms, molecules and every other form of matter in the universe. This includes us, our planet and, of course, the Large Hadron Collider. In that sense, everything since the Big Bang has been one giant quantum experiment, in which all the particles in the universe, including those we think of as making up the Earth and our own bodies, are involved. But if theory tells us we're among the sets of particles involved a giant quantum experiment, the position I've just outlined tells us we can't justify any statement about what has happened or is happening until the experiment is over. Only at the end, when we might perhaps imagine some technologically advanced alien experimenters in the future looking at the final state of the universe, can any meaningful statement be made.

Of course, this final observation will never happen. By definition, no one is sitting outside the universe waiting to observe the final outcome at the end of time. And even if the idea of observers waiting outside the universe made sense – which it doesn't – on this view their final observations still wouldn't allow them to say anything about what happened *between* the Big Bang and the end of time. We end up concluding that quantum theory doesn't allow us to justify making any scientific statement at all about the past, present or future. Our most fundamental scientific theory turns out to be a threat to the whole enterprise of science. For these and related reasons, the Copenhagen interpretation gradually fell out of general favour.

Its great rival was first set out in a 1957 paper [7] and Princeton PhD thesis written by one of the stranger figures in the history of 20th-century physics, Hugh Everett III. Rather unromantically, and very unusually for a highly original thinker and talented physicist, Everett abandoned theoretical physics after he had published his big idea. A good deal of his subsequent career was spent in military consultancy, advising the US on strategies for fighting and 'winning' a nuclear war against the USSR, and the bleakness of this chosen path presumably contributed to his chain-smoking, alcoholism and depression. Everett died of a heart attack at the age of 51; possibly we can infer something of his own ultimate assessment of his life's worth from the fact that he instructed his wife to throw his ashes in the trash. And yet, despite his detachment from academic life (some might say from all of life), Everett's PhD work eventually became enormously influential.

One way of thinking about his ideas on quantum theory is that our difficulties in getting a description of quantum reality arise from a tension between the mathematics – which, as we have seen, tells us to make calculations involving many different possible stories about what might have really happened – and the apparently incontrovertible fact that, at the end of an experiment, we see that only one thing actually did happen. This led Everett to ask a question that seems at first sight stupid, but which turns out to be very deep: how do we know that we only get one outcome to a quantum experiment? What if we take the hint from the mathematics and consider a picture of reality in which many different things actually do happen – everything, in fact, that quantum theory allows? And what if we take this to its logical conclusion and accept the same view of cosmology, so that all the different possible histories of the evolution of the universe are realised? We end up, Everett argued, with what became known as a 'many worlds' picture of reality, one in which it is constantly forming new branches describing alternative – but equally real – future continuations of the same present state.

On this view, every time any of us does a quantum experiment with several possible outcomes, all those outcomes are enacted in different branches of reality, each of which contains a copy of our self whose memories are identical up to the start of experiment, but each of whom sees different results. None of these future selves has any special claim to be the real one. They are all equally real – genuine but distinct successors of the person who started the experiment. The same picture holds true more generally in cosmology: alongside the reality we currently habit, there are many others in which the history of the universe and our planet was ever so slightly different, many more in which humanity exists on Earth but the course of human history was significantly different from ours, and many more still in which nothing resembling Earth or its inhabitants can be found.

This might sound like unbelievable science fiction. To such a gibe, Everett and his followers would reply that science has taught us many things that seemed incredible at first. Other critics object that the 'many worlds' scenario seems like an absurdly extravagant and inelegant hypothesis. Trying to explain the appearance of one visible reality by positing an infinite collection of invisible ones might seem the most deserving candidate in the history of science for a sharp encounter with Occam's razor. But to this, too, Everettians have an answer: given the mathematics of quantum theory, on which everyone agrees, their proposal is actually the *simplest* option. The many worlds are there in the equations. To eliminate them you have to add something new, or else change them – and we don't have any experimental evidence telling us that something should be added or that the equations need changing.

Everettians might have a point, then, when they argue that their ideas deserve a hearing. The problem is that, from Everett and his early followers onwards, they have never managed to agree on a clear story about how exactly this picture of branching worlds is supposed to emerge from the fundamental equations of quantum theory, and how this single world that we see, with experimental outcomes that are apparently random but which follow definite statistical laws, might then be explained. One of the blackly funny revelations in Peter Byrne's biography *The Many Worlds of Hugh Everett III* (2010) [8] was the discovery of Everett's personal copy of the classic text *The Many-Worlds Interpretation of Quantum Mechanics*[9], put together in 1973 by the distinguished American physicist Bryce DeWitt and a few of Everett's other early supporters. To DeWitt's mild criticism that 'Everett's original derivation [of probabilities]… is rather too brief to be entirely satisfying', Everett scribbled in the margins 'Only to you!' and 'Goddamit [sic] you don't see it'. On another paper addressing the same issue, his comment was the single word 'bullshit'. Although generally in more civil terms, Everettians have continued to argue over this and related points ever since.

Indeed, the big unresolved, and seemingly unsolvable, problem here is how statistical laws can possibly emerge at all when the Everettian meta-picture of branching worlds has no randomness in it. If we do an experiment with an uncertain outcome, Everett's proposal says that everything that could possibly happen (including the very unlikely outcomes) will in fact take place. It's possible that Everettians can sketch some explanation of why it seems to 'us' (really, to any one of our many future successors) that 'we' see only one outcome. But that only replaces 'everything will actually happen' with 'anything could seem to happen to us' – which is still neither a quantitative nor a falsifiable scientific statement. To do science, we need to able to test statements such as 'there's a one-in-three chance X will happen to us' and 'it's incredibly unlikely that Y will happen to us' – but it isn't at all obvious that Everett's ideas support any such statements.

Everettians continue to devote much ingenuity to deriving statements involving probabilities from the underlying deterministic many-worlds picture. One idea lately advocated by David Deutsch and David Wallace of the University of Oxford is to try to use decision theory, the area of mathematics that concerns rational decision-making, to explain how rational people should behave if they believe they are in a branching universe. Deutsch and Wallace start from a few purportedly simple and natural technical assumptions about the preferences one should have in a branching world and then claim to show that rational Everettians should behave *as though* they were in an uncertain probabilistic world following the statistical laws of quantum theory, even though they believe their true situation is very different.

One problem with this line of thought is that the assumptions turn out not to seem especially natural, or even properly defined, on close inspection. The easiest way to understand this is to look for rationally defensible strategies for life in a branching universe other than the ones Deutsch and Wallace advocate. One example I rather like (because it makes the point succinctly, not because it seems morally attractive) is that of future self elitism, which counsels us to focus only on the welfare of our most fortunate and successful future successor, perhaps on the premise that our best possible future self is our truest self. Future self elitists don't worry about the odds of a particular bet, only about the best possible payoff. Thus they violate Deutsch and Wallace's axioms, but it is hard to see any purely logical argument against their decisions.

Another issue is that, as several critics have pointed out, whatever one thinks of Deutsch and Wallace's proposed rational strategy, it answers a subtly different question to the one that Everettians were supposed to be addressing. The question 'What bets should I be happy to place on the outcomes of a given experiment, given that I believe in Everettian many-worlds?' is certainly a question that relates something we normally try to answer using probabilities with the many-worlds picture. In that sense, it makes some sort of connection between probabilities and many worlds – and since we've seen how hard that is to achieve, it's easy to understand why Everettians (at least initially) are enthusiastic about this accomplishment. But, unfortunately, it's not the sort of connection we need. The key scientific question is why the experimental evidence for quantum theory justifies a belief in many worlds in the first place. Many Everettians – from Everett and DeWitt onwards – have tried to give a satisfactory answer to this. Many critics (myself included) appreciate the cunning of their attempts but think they have all failed.

If we cannot get a coherent story about physical reality from the Copenhagen interpretation of quantum theory and we cannot get a scientifically adequate one from many-worlds theory, where do we turn? We could, as some physicists suggest, simply give up on the hope of finding any description of an objective external reality. But it is very hard to see how to do this without also giving up on science. The hypothesis that our universe began from something like a Big Bang, our account of the evolution of galaxies and stars, the formation of the elements and of planets and all of chemistry, biology, physics, archaeology, palaeontology and indeed human history – all rely on propositions about real observer-independent facts and events. Once we assume the existence of an external world that changes over time, these interrelated propositions form a logically coherent set; chemistry depends on cosmology, evolution on chemistry, history on evolution and so on. Without that assumption, it is very hard to see how one might make sense of any of these disciplines, let alone see a unifying picture that underlies them all and explains their deep interrelations and mutual dependence.

If we can't allow the statement that dinosaurs really walked the Earth, what meaningful content could biology, palaeontology or Darwinian evolution actually have? It's even harder to understand why the statement seems to give such a concise explanation of many things we've noticed about the world, from the fossil record to (we think) the present existence of birds, if it's actually just a meaningless fiction. Similarly, if we can't say that water molecules really contain one oxygen and two hydrogen atoms – or at least that something about reality that supports this model – then what, if anything, is chemistry telling us?

Physics poses many puzzles, and the focus of the physics community shifts over time. Most theoretical physicists today do not work on this question about what really happens in quantum experiments. Among those who think about it at all, many hope that we can find a way of thinking about quantum theory in which reality somehow evaporates or never arises. That seems like wishful thinking to me.

The alternative, as John Bell recognised earlier and more clearly than almost all of his contemporaries, is to accept that quantum theory cannot be a complete fundamental theory of nature. (As mentioned above, Einstein also believed this, though at least partly because of arguments that Bell was instrumental in refuting.)

Bell was one of the last century's deepest thinkers about science. As he put it, quantum theory 'carries in itself the seeds of its own destruction': it undermines the account of reality that it needs in order to make any sense as a physical theory. On this view, which was once as close to heresy as a scientific argument can be but is now widely held among scientists who work on the foundations of physics, the reality problem is just not solvable within quantum theory as it stands. And so, along with the variables that describe potentialities and possibilities, we need to supplement our quantum equations with quantities that correspond directly to real events or things – real 'stuff' in the world.

Bell coined the term *beables* to refer to these elusive missing ingredients. 'Beable' is an ugly word but a useful concept. It denotes variables that are able to 'be' in the world – hence the name. And indeed it turns out that we *can* extend quantum theory to include beables that would directly describe the sort of reality we actually see. Some of the most interesting work in fundamental physics in the past few decades has been in the search for new theories that agree with quantum theory in its predictions to date, but which include a beable description of reality, and so give us a profoundly different fundamental picture of the world.

What sort of quantities might do the trick? One early idea comes from Louis de Broglie, whom we met earlier, and David Bohm, an American theoretical physicist who fled

McCarthyite persecution and spent most of his career at the University of London. The essence of their proposal is that, in addition to the mathematical quantities given to us by quantum theory, we also have equations defining a definite path through space and time for each elementary particle in nature. These paths are determined by the initial state of the universe and, in this sense, de Broglie-Bohm theory [4] can be thought of as a deterministic theory, rather like the pre-quantum theories given by Newton's and Maxwell's equations. Unfortunately, de Broglie and Bohm's equations also share another property of Newton's equations: an action at any point in space has instantaneous effects on particles at arbitrarily distant points.

Because these effects would not be directly detectable, this would not actually allow us to send signals faster than light, and so it does not lead to observations that contradict Einstein's special theory of relativity. It does, however, very much violate its spirit, as well as the beautiful symmetry principles incorporated in the underlying mathematics. For this reason, and also because de Broglie and Bohm's ideas work well for particles but are hard to generalise to electromagnetic and other fields, it seems impossible to find a version of the scheme that is consistent with much of modern theoretical physics. Still, de Broglie and Bohm's great achievement was to show that we *can* find a mathematically consistent description of reality alongside quantum theory. When it first emerged, their work was largely unappreciated, but it led to many of Bell's insights into the quantum reality problem and blazed a trail for later theorists.

In the 1980s, a much more promising avenue opened up, thanks to the efforts of Giancarlo Ghirardi, Alberto Rimini, Tullio Weber and Philip Pearle [6,7], three European theorists and an American. Their approach became known as the 'spontaneous collapse' model and their brilliant insight was that we can find mathematical laws that describe how the innumerable possible outcomes encoded in a quantum description of an experiment get reduced to the one actual result that we see. As we have already noted, the tension between these two descriptions is at the heart of the quantum reality problem.

When using standard quantum theory, physicists often say that the wave function – a mathematical object that encodes all the potential possibilities – 'collapses' to the measured outcome at the end of an experiment. This 'collapse', though, is no more than a figure of speech, which only highlights the awkward fact that we do not understand what is really happening. By contrast, in Ghirardi-Rimini-Weber-Pearle models, collapse becomes a well-defined mathematical and physical process, taking place at definite points in space, following precise equations and going on all the time in the world around us, whether or not we are making measurements. According to these new equations, the more particles there are in a physical system, the faster the collapse rate. Left isolated, a single electron will collapse so rarely that we essentially never see any effect. On the other hand, anything large enough to be visible – even a dust grain – has enough particles in it that it collapses very quickly compared to human perception times. (In Schrödinger's famous thought experiment, the cat's quantum state would resolve in next to no time, leaving us with either a live cat or a dead one, not some strange quantum combination of both.)

One way of thinking about reality in these models, first suggested by Bell [10], is to take the beables to be the points in space and time at which the collapses take place. On this view, a dust grain is actually a little galaxy of collapse points, winking instantaneously in and out of existence within or near to (what we normally think of as) the small region of space that it occupies. Everything else we see around us, including our selves, has the same sort of pointillistic character.

Collapse models do not make exactly the same predictions as quantum theory, which could turn out to be either a strength or a weakness. Since quantum theory is very well confirmed, this disagreement might seem to rule these new models out. However, the exact rate of collapses per particle is a free parameter that is not fixed by the mathematics of the basic proposal. It is perfectly possible to tailor this value such that the differences between collapse model predictions and those of quantum theory are so tiny that no experiment to date would have detected it, and at the same time large enough that the models give a satisfactory solution to the reality problem (ie, everything that seems definite and real to us actually is real and definite).

That said, we presently have no theoretically good reason why the parameter should be in the range that allows this explanation to work. It might seem a little conspiratorial of nature to give us the impression that quantum theory is correct, while tuning the equations so that the crucial features that give rise to a definite physical reality are – with present technology – essentially undetectable. On the other hand, history tells us that deep physical insights, not least quantum theory itself, have often come to light only when technology advances sufficiently. The first evidence for what turns out to be a revolutionary change in our understanding of nature can often be a tiny difference between what current theory predicts and what is observed in some crucial experiment.

There are other theoretical problems with collapse models. Although they do not seem to conflict with special relativity or with field theories in the way that de Broglie-Bohm theory does, incorporating the collapse idea into these fundamental theories nevertheless poses formidable technical problems. Even on an optimistic view, the results in this direction to date represent work in progress rather than a fully satisfactory solution. Another worry for theorists in a subject where elegance seems to be a surprisingly strong indicator of physical relevance is that the mathematics of collapse seems a little ad hoc and utilitarian. To be fair, it is considerably less ugly than the de Broglie-Bohm theories, which to a purist's eye more closely resemble a Heath Robinson contraption than the elegant machinery we have come to expect of the laws of physics. But compared with the extraordinary depth and beauty of Einstein's general theory of relativity, or of quantum theory itself, collapse models disappoint.

This could simply mean that we have not properly understood them, or not yet seen the majestic deeper theory of which they form a part. It seems likelier, though, that collapse models are at best only a step in roughly the right direction. I suspect that, like de Broglie-Bohm theory, they will eventually be seen as pointers on the way to a deeper understanding of physical reality – extraordinarily important achievements, but not fundamentally correct descriptions.

There is, however, one important lesson that we can already credit to collapse models. De Broglie-Bohm theory suffers from the weakness that its experimental predictions are *precisely* the same as those of quantum theory, unlike collapse models that, as we have noted, are at least in principle testably different. The beables in de Broglie-Bohm theory – the particle paths – play a rather subordinate role: their behaviour is governed by the wave function that characterises all the possible realities from which any given set of paths is drawn, but they have no effect on that wave function. In metaphysical language, the de

Broglie-Bohm theory beables are *epiphenomena*. The American psychologist William James once critically and poetically described [11] the English biologist T H Huxley's view of human consciousness as 'Inert, uninfluential, a simple passenger in the voyage of life, it is allowed to remain on board, but not to touch the helm or handle the rigging'. Much the same might be said of a de Broglie-Bohm beable. Collapse-model beables, on the other hand, give as good as they get. Their appearance is governed by rules involving the quantum wave function, and yet, once they appear, they in turn alter the wave function. This makes for a far more interesting theory, mathematically as well as scientifically.

It's tempting to declare this as a requirement for any variable in a fundamental theory of physics – or at least, any variable that plays as important a role as the beables are meant to play: it should be mathematically active, not purely passive. Any interesting solution to the quantum reality problem should (like collapse models but unlike de Broglie-Bohm theory) make experimentally testable predictions that allow us to check our new description of reality.

How might we do that? Assuming these ideas are not entirely wrong, what sort of experiments might give us evidence of a deeper theory underlying quantum theory and a better understanding of physical reality? The best answer we can give at present, if collapse models and other recent ideas for beable theories are any guide, is that we should expect to see something new when some relevant quantity in the experiment gets large. In particular, the peculiar and intriguing phenomenon called quantum interference – which seems to give direct evidence that different possible paths which could have been followed during an experiment all contribute to the outcome – should start to break down as we try to demonstrate it for larger and larger objects, or over larger and larger scales.

This makes some intuitive sense. Quantum theory was developed to explain the behaviour of atoms and other small systems, and has been well tested only on small scales. It would always have been a brave and perhaps foolhardy extrapolation to assume that it works on all scales, up to and including the entire universe, even if this involved no conceptual problems. Given the self-contradictions involved in the extrapolation and the profound obstacles that seem to prevent any solution of the reality problem within standard quantum theory, the most natural assumption is that, like every previous theory of physics, quantum mechanics will turn out only approximately true, applying within a limited domain only.

A number of experimental groups around the world are now trying to find the boundaries of that domain, testing quantum interference for larger and larger molecules (the current record is for molecules comprising around 1,000 atoms), and ultimately for small crystals and even viruses and other living organisms. This would also allow us to investigate the outlandish but not utterly inconceivable hunch that the boundaries of quantum theory have to do with the complexity of a system, or even with life itself, rather than just size. Researchers have proposed space-based experiments to test the interference between very widely separated beams and will no doubt spring into action once quantum technology becomes available on satellites, as it probably will in the next few years.

With luck, if the ideas I have outlined are on the right lines, we might have a good chance of detecting the limits of quantum theory in the next decade or two. At the same time we can hope for some insight into the nature and structure of physical reality. Anyone who expects it to look like Newtonian billiard-balls bouncing around in space and time, or anything remotely akin to pre-quantum physical ideas, will surely be disappointed. Quantum theory might not be fundamentally correct, but it would not have worked so well for so long if its strange and beautiful mathematics did not form an important part of the deep structure of nature. Whatever underlies it might well seem weirder still, more remote from everyday human intuitions, and perhaps even richer and more challenging mathematically. To borrow a phrase from John Bell, trying to speculate further would only be to share my confusion. No one in 1899 could have dreamed of anything like quantum theory as a fundamental description of physics: we would never have arrived at quantum theory without compelling hints from a wide range of experiments.

The best present ideas for addressing the quantum reality problem are at least as crude and problematic as Bohr's model of the atom. Nature is far richer than our imaginations, and we will almost certainly need new experimental data to take our understanding of quantum reality further. If the past is any guide, it should be an extraordinarily interesting scientific journey.